\documentclass[11pt,titlepage,a4paper]{article}
\usepackage{Bozho-LaTeX-Style}	
\usepackage{Bozho-LaTeX-Logo}	

\usepackage{cite}	
\usepackage{tabularx}	
\usepackage{varioref}	

\title{\bfseries	\vspace*{-1.678902345in}
{\huge Measuring\\ the evaluation and impact\\ of scientific works and
their authors}
}

\vspace{1.7ex}

\author{
Bozhidar Z.\ Iliev
\thanks{Laboratory of Mathematical Modeling in Physics,
Institute for Nuclear Research and \mbox{Nuclear} Energy,
Bulgarian Academy of Sciences,
Boul.\ Tzarigradsko chauss\'ee~72, 1784 Sofia, Bulgaria}
\thanks{E-mail address: bozho@inrne.bas.bg}
\thanks{URL: http://theo.inrne.bas.bg/$\sim$bozho/}
}

\date{
 \vspace{2.27ex}\ShortTitle{Measuring the impact of scientific
works and their authors}	\\[0.27ex]
 \vspace{3.27ex}
\small
	\begin{tabular}{r@{$\colon\to~$}l}
	\end{tabular} \\[1.27ex]
 \small
	\begin{tabular}{r@{$\colon~$}l}
 \normalsize\sffamily\bfseries
	\end{tabular} \\[-0.27ex]
\normalsize
 \vspace{4.27ex}{\Huge\BOZHO}	\\[4.27ex]
 \vspace{0.27ex}\KeyWords{
Impact of published work(s),
impact of author(s),
evaluation of author(s),
author achievements,
citation,
citations,
implicit citations,
citations analysis,
\\
bibliometrics,
bibliometric indicators,
bibliometric metrics,
\\
bibliometric indices,
Hirsch index,
\\
peers, peer judgment, peer review
}	\\[0.27ex] }

\begin{document}		

\renewcommand{\thepage}{\roman{page}}

\renewcommand{\thefootnote}{\fnsymbol{footnote}} 
\maketitle				
\renewcommand{\thefootnote}{\arabic{footnote}}   

\tableofcontents		


\begin{abstract}

	The work is partially a review article and partially a research paper.
Problems for evaluation and impact of published scientific works and their
authors are discussed at theoretical level. The role of citations in this
process is pointed out. Different bibliometric indicators are reviewed in this
connection and ways for generation of new bibliometric indices are given. The
influence of different circumstances, like self\ndash citations, number of
authors, time dependence and publication types, on the evaluation and impact of
scientific papers are considered. The repercussion of works citations and their
content is investigated in this respect. Attention is paid also on implicit
citations which are not covered by the modern bibliometrics but often are
reflected in the peer reviews. Some aspects of the Web analogues of citations
and new possibilities of the Internet resources in evaluating authors
achievements are presented.

\end{abstract}

\renewcommand{\thepage}{\arabic{page}}


\section {Introduction}
\label{Introduction}

	Can the scientific output of a scientists be measured quantitatively?
We often said that someone has better achievements than other person but explain
this with non-strict words and opinions of experts in the corresponding
field of research which certainly can't be measured quantitatively (except
counting some kind of votes via a qualitative procedure of rating). To such
qualifications often are added strict number measures like the number of
published papers and their (known) number of total citations. The former is a
measure of author productivity while the latter one is considered as his/her
impact on (other) authors. Just here comes into action the bibliometrics~%
\footnote{~%
Sometimes the bibliometrics is called scientometrics but our opinion is that
these are different things~\cite{Glanzel-2010}.%
}
which has as input data the raw information about an author published (and
publicly available) works and their recorded influence on other published works
and as an output gives quantitative conclusions concerning the author.~%
\footnote{~%
Here and below we talk about author(s) but in the most cases the text is
true for group of authors, journal, university, county \etc%
}
This process is well described in~\cite{Pendlebury-2010}.

	The bibliometrics provides a number of already established numerical
characteristics of authors publications and their
citations~\cite{Rehn_et_al-2007, Durieux_Gevenois-2010}, known as bibliometric
indicators, such as number of publications (total and for some period of time),
number of publications in top journals, number of citations (total and for some
period of time), citations per publication, top 5\%\ citations, \etc
	Starting from 2005 Hirsch paper~\cite{Hirsch-2005} there were introduced a
number of new bibliometric indicators~\cite{Alonso_et_al-2009} like the Hirsch
index~$h$ and different (Hirsch\ndash like) indices that modify it in ways that
compensate some its disadvantages.
	Regardless of these rigorous measures, the peer judgements remain
leading in takeing decisions about the achievements of papers and their authors.
On statistical level is observed a correlation between assessment by different
bibliometric indicators and quality judgment of
peers~\cite{h-index_for_147_chemistry_research_groups, Waltman_et_al-2011,
Bornmann-2011}. This naturally suggest the both methods to be used as
complimentary to each other.

	This paper has aspects of a review article and a research
paper simultaneously. Sect.~\ref{Sect2} points to some peculiarities of citing in
different types of publications and concerns the problem of self\ndash
citations. Sect.~\ref{Sect3} is devoted to citations lists
and ways for preparing them. Different forms of citation lists are presented in
Sect.~\ref{Sect4}. Special attention is paid to citations of works with more
than one author and to citing papers with multiple authors. Sect.~\ref{Sect5}
deals with some bibliometric indices. The Hirsch index, certain its
modifications and complimentary to it indices are recalled. Ways for generation
of new bibliometric indices are provided. Sect.~\ref{Sect6} concerns problems
like self\ndash citation, number of authors, highly/low cited papers, and time
dependence of the citations. Connections between citations and scientific
achievements are discussed in Sect.~\ref{Sect7}. In Sect.~\ref{Sect8} are
presented some aspects of the problem on how the content of a paper may
influence its evaluation and impact. The implicit citations are discussed in
Sect.~\ref{Sect9}. The role of the peers is mentioned in Sect.~\ref{Sect10}.
The paper ends with a final discussion in Sect.~\ref{Conclusion}.

	As the author of this paper works mainly in the field of
(mathematical) physics and mathematics, the problems investigated in it
concern physics literature but it is likely that they apply also to other
publishing fields.


\section {When a work is cited?}
\label{Sect2}

	In physics any scientist builds his/her work on the base of earlier
existing works and for this reason new works/publications cite the works they
are build upon; a deep analysis of this process is contained
in~\cite{Moed-2005} and more particular reasons for citing are presented
in~\cite[Sect~4.1]{Glanzel-2003}. In this way is made a link with already
existing knowledge and is paid tribute to the work of the scientist that have
contributed to it. In this sense the citation is part of the process of linking
of a work with the knowledge preceding it.
	It is known that the more citations a published paper has, the more
impact it has on the other authors~\cite{Nosek-2010} but the problem is to
evaluate this impact quantitatively. As a consequence of this the (number of)
citations of a paper is a measure for its impact on other works and scientists.
Respectively, the (number of) citations of the papers of an author is a measure
for his/her impact.

	The reasons for citing a particular work and putting it in a
reference/bibliography list of papers are numerous and depend on the type of
the publication in which it is cited, its content and the authors. Below we
shall try to analyze this problem for some kinds of works and to make
conclusions which may be useful for finding criterions for evaluation of
publishing (and, possibly, scientific) activity of a person. A comprehensive
analysis of the reasons for citing can be found in~\cite{Moed-2005}.
In~\cite[pp.~11--13]{Rehn_and_Kronman-2008} is presented a list of factors that
affect the number of citations.

	Besides the types of works considered below, there may be
distinguished many other types of published works. Moreover, there exist works
that are of mixed type, \eg a handbook or review paper containing new results
and thus having elements of a research work. Here we are not going to present a
''complete'' list and analysis of publication types and note that an honest
citation of a paper is intended to point readers attention to it and may mean
that the author(s) has (have) used some information from the cited work.

\subsection {Citation in a research paper}
\label{Subsect2.1}

	The research papers are regular discovery accounts and are usually
in the form of journal articles, preprints, electronic preprints and others. As
their short versions can be regarded the meeting communications (abstracts,
full or part text articles), short communications, letters (possibly to
editors), notes, corrections/additions of/to earlier works, \etc At present a
typical (full text) research paper has an introduction, main body and concluding
part.

	Some of the roles of the introduction are: (i)~to present the main
problems that will be investigated further in the work; (ii)~to pay attention
to (some of) the existing results on them; (iii)~to fix certain notations,
concepts and results that will be used in the work; (iv)~to point to the
history and possible future developments of the items of the work. So, citing a
paper in the introductory part of a research work may mean different things
like:
    \begin{enumerate}
\item
	It belongs to a general list of references on some item considered in
the work.
\item
	It contains essential results that will be used, developed, commented,
\etc in the work.
\item
	Contains problem(s) that will be investigated in the citing work.
\item
	It is of pure historical interest; \eg representing a wrong theory.
    \end{enumerate}

	The main body of a research work contains rigorous statement of some
problems, their analysis and, possibly, their solutions. Respectively, normally
a paper is cited here when it is directly connected with these problems and its
content is (partially) used in the work. This meas that a paper cited in the
main body has, generally, more impact on the work than a paper cited in the
introduction (if something else is not stated explicitly).

	At last, the purpose of the conclusion may be: (i)~summarizing the
outcome of the main body of the work, (ii)~comments/analysis of the results
obtained; (iii)~making connections with other works containing results of
interests; (iv)~pointing to non-solved problems and further developments.
Correspondingly, here typically a paper is cited when it poses similar problems
but its results do not influence directly the main developments of the work.



\subsection {Citation in a review work}
\label{Subsect2.2}

	The main aim of a review work is to bring together results obtained in
research papers for some period of time. However, the particular realization of
such a work may be done in quite different ways, for example:
    \begin{enumerate}
\item
	A simple list of literature with possible comments.
\item
	An independent presentation of the material; \eg in a book or book-like
paper.
\item
	An unified presentation of groups of papers in different sections
forming the main body of the work.
    \end{enumerate}

In any case, a review paper is generally not suppose to contain new results.
Its main purpose is to put in a single place results that can be found in
different sources which form the main part of the citation list of such kind of
a work. In this sense, most of the papers cited in a review work are
essentially used. Besides, a citation of a particular part of a review work may
be considered, in some sense, as citation of the original papers on which the
cited part rest.



\subsection {Citation in a handbook, encyclopedia and similar works}
\label{Subsect2.3}

	The handbooks and encyclopedias may be regarded as review works but
they have more specific structure, presentation and usually cover larger arias
of materials. A typical work of this kind consists of series of (alphabetically
ordered) separate papers (articles) with possible cross\ndash references between
them. They contain normally only presentation of facts (results, theorems,
methods) with little comments and their reference lists are restricted to
represent (details on) these facts. So, any paper cited in a handbook or
encyclopedia is essentially used in it. Besides, citing an article of a such a
work may be regarded as an indirect citation (of some) of the papers in its
reference list.



\subsection {Citation in textbook}
\label{Subsect2.4}

The purpose of a textbook is learning the material presented in it. This
usually limits the citations in it, if any, to publications that are: (i)~other
textbooks on the same or similar material; (ii)~containing original (\eg
historical) material on the covered items; (iii)~further developments on the
subject(s) covered; (iv)~used by the author(s) to write it.



\subsection {Self-citations}
\label{Subsect2.5}

	There are many reasons when an author cites his/her own paper(s).
Normally this is done when the author has previous publications on the
subject(s) of the work where self\ndash citations appear and  he/she finds them
essential in the context where they are cited. In this sense, the self\ndash
citations reveal the self\ndash impact of an author and should be treated on the
same footing as any other citations.

	It should be said that there are authors that intentional cite their
own papers for, let us say, ''non\ndash scientific'' reasons; \eg popularizing
own works, extending the list of citations of their works \etc The author of
these lines would like to think that these are exceptional cases, at least in
the case of research papers and may be neglected in the general case. However,
if there are facts that a particular author belongs to this category of
authors, then he/she may be blamed as non\ndash hones with respect to his/her
citation list and the self\ndash citations in it should be considered
critically or neglected at all.

	We shall return on the problem for self-citations in
Subsection~\ref{Subsect6.1}.



\subsection {Inferences}
\label{Subsect2.6}

	Without considering other types of publications and treating
self\ndash citations as ordinary citations, we may point to some of the main
reasons for citations:
    \begin{enumerate}
\item
	Using particular information, like results, methods and formulae, form
the cited works.
\item
	Pointing to texts from the cited work without using them.
\item
	Pointing the readers attention to works connected to the subject(s)
considered in the citing work.
\item
	Presenting list(s) of publications on some item(s)  .
    \end{enumerate}

	The impact of a cited paper on the citing one depends on the category
to which it belongs. It seems that most weight should be given to citing paper
from the first of the above category. However, it is unlikely that particular
numerical weights can be assigned to some or all categories of the citations
and, as a result of this, the arrangement of these categories by weights is
qualitative. Of course, the impact of a paper depends on its content and the
contents of the works citing it.



\section {Lists of citations}
\label{Sect3}

	Nowadays there is an understanding that the more citations an author
has, the greater is his/her impact in Science.~%
\footnote{~%
Here is excluded the problem of the content of the papers cited as well as the
context in which the citations are made. For instance, an evident counter
example of this understanding is a citation in which is pointed plagiarism in
the cited work.
}
For this and other reasons a lot of authors make lists of works citing their
own papers. Such a list may have different purposes like:
    \begin{itemize}
\item
	To show other scientists how his/her works are used by other authors.
\item
	It is needed for some official (possibly internal) account.
\item
	It may be a part of the reasons for obtaining scientific degree or a
promotion.
\item
	It may be a reason for authors proud or simply a way to tell other
scientist which authors have used his/her works.
    \end{itemize}

	A preparation of author citation list is not an easy task in times
when there are literally tens of thousands of scientific journals,
institutional/university annual reports, books \etc published \eg monthly or
annually. The easiest way to make such a list is via the Internet based
databases like:
    \begin{enumerate}
\item
	Google Scholar (free) with URL http://scholar.google.com/.
\item
	Web of Science~%
\footnote{%
The Web of Science (WoS) is an electronic version of the Science Citation Index
(SCI)~\cite{Garfield-2007}.%
}
(paid) with URL~http://thomsonreuters.com/products\_services/science/ science\_products/a-z/web\_of\_science/.
\item
	Science Direct (paid) with URL http://www.sciencedirect.com/.

\item
	SCOPUS (paid) with URL http://www.scopus.com.
\item
	CiteSeerX (free) with URL http://citeseerx.ist.psu.edu/. It has
replaced the database CiteSeer.
\item	Microsoft Academic Search (free) with URL
http://academic.research.microsoft.com/

    \end{enumerate}
	The above databases cover differently
different scientific fields and types of
publications~\cite[pp.~349--350]{Durieux_Gevenois-2010} like journal articles,
electronic preprints, books/monographs, conference reports, theses, \etc A
concise and good analysis of them is given in~\cite{Kermarrec_et_al-2007}.
	In general, they give overlapping but not identical
results~\cite{Bibliometrics-an_introduction, Bar-Ilan-2008, Kumar-2009,
Alonso_et_al-2009}. A description of some advantages and disadvantages of
Google Scholar and Thomson ISI web of science is given in~\cite{Harzing-2010}.

	A less efficient way for finding citations is to search the Web for
some combinations of key\ndash words including the name(s) of the author whose
citations are looked for and possibly the names of the authors who may cite
him/her.

	For preparation of citation lists in the field of physics and/or
mathematics one can use also the sites:
    \begin{enumerate}
\item
	arXiv with URL http://arXiv.org.
\item  IOP eprint web with URL http://eprintweb.org which is based on the
arXiv.
\item
	 SAO/NASA Astrophysics Data System (ADS) with URL
http://adsabs.harvard.edu
 \item
        INSPIRE with URL http://inspirehep.net.
    \end{enumerate}

	Of course, for making
citation lists one may use more ''conventional'' resources like
    \begin{enumerate}
\item
	(accidental) reading of scientific papers.
\item
	 personal acquaintance with scientists.
\item
	consultations with the Science Citation Index (SCI) of the Thomson
Reuters Institute for Scientific Information which is a paper version of The
Web of Science (WoS).

    \end{enumerate}

	It is important to note that the data in a citation list should be
publicly available as otherwise it is (almost) impossible to check/verify
independently its trueness.

	The completeness of a citation list depends on the sources used, \ie
the data sets from which it is prepared. In this sense, a particular citation
list gives also a lower limit on the number of works with non-zero citations as
well the number of their citations.

	For the purposes of this paper we assume below that an author
citation list includes all his/her published papers; in particular, these with
zero number of citations.


\section {Analysis and forms of citation lists}
\label{Sect4}

	To make conclusions based on citations from a list of citations of an
author it is convenient to arrange the author's papers in order of descending
number of their citations. If some works have equal number of citations, then
their relative order is insignificant and they can be arrange in such a list in
an arbitrary way relative to each other, \eg alphabetically by their titles.
The consecutive number of a paper in such a list is called its rank (in this
list). So, at this stage, a citation list of an author with $n\ge1$ published
papers can be represented as like the Table~\vref{Table_4.1}.

\renewcommand{\arraystretch}{1.1}	

\begin{table}[ht!]

\vspace{1ex}
\centering
	\begin{tabular}
		{@{} | >{$}c<{$} | >{$}c<{$} | >{$}c<{$} | @{}}
\hline
\text{Number}	    & \text{Rank} & \text{Paper} \\
\text{of citations} &		  & \text{description}
\\ \hline
c_1 & 1  & p_1   \\
c_2 & 2  & p_2   \\
\vdots & \vdots & \vdots \\
c_n & n  & p_n   \\ \hline
	\end{tabular}
\noindent
   \begin{minipage}{0.85\textwidth}
\caption{
\label{Table_4.1}
Initial example form of a citation list.
\newline \small
Here $c_i,\ i=1,\dots,n$, is the number of citation of the paper with rank $i$
and description $p_i$. By definition $c_i\ge c_{i+1}$ for $i=1,\dots,n-1$ and
it is possible that $c_{i}=0$ for $i\ge n_0$ for some $n_0\in\{1,\dots,n\}$.
}
   \end{minipage}
 \end{table}

 	A little information can be obtained form
Table~\vref{Table_4.1} without a comparison
 with similar tables for other authors. The main inference is that the more a
paper is closer to the top of the list, the more it has been used by the
authors and vice versa, the closer a paper is to the table end, the less it
has been used. At this stage, the paper rank is a measure of its importance for
the authors: the lass the rank, the more important a paper is and \vv  As a
quantitative measure for this opinion may serve the numbers
    \begin{equation}    \label{4.1}
c_i^r := \frac{c_i}{\sum_{i=1}^nc_i}
    \end{equation}
which are the citation numbers normalized by their sum, so that
$0\le c_i^r\le 1$ and $\sum_{i=1}^n c_i^r = 1$. Of course, here we suppose that
the author has at least one published work with least one citation.

	Usually, there is a number $n_0\in\{1,\dots,n\}$ such that $c_{i}=0$
for $i\ge n_0$, \ie the papers with rank greater then or equal to $n_0$ have no
citations and the first $n_0-1$ papers in the table (with rank less than $n_0$)
have at least one citation. If such a number $n_0$ exists, the ratio
    \begin{equation}    \label{4.2}
E := \frac{n_0-1}{n}
    \end{equation}
can be called author effectiveness (or coefficient of performance (COP) or
coefficient of efficiency) as it measures how much of his/her published works
have been used by (other) authors. If there is no a number $n_0$ with the
properties required, we set $n_0=n+1$ and $E=1$. So that $0\le E \le 1$.

	Obviously, the greater the author efficiency, the more of his/her
published works have been used by authors and possibly influenced their papers.

\subsection{Cited papers with multiple authors}
\label{Subsect4.1}

	Till this point we have not mentioned problems concerning the number
of authors of any particular work in which the author has contributed (as a
coauthor). Since we aim to make conclusions concerning a particular person, the
above written is valid in a case when all papers in table~\vref{Table_4.1} are
written by a single person, \ie there are not other co-authors. However, in the
general case, when the paper $p_i$ has $a_i\ge 1$ authors, the needed for our
purposes modification of table~\vref{Table_4.1} may look like the next
table~\vref{Table_4.2}.

\renewcommand{\arraystretch}{1.1}	

\begin{table}[ht!]


\vspace{1ex}
\centering
	\begin{tabular}
		{@{} | >{$}c<{$} | >{$}c<{$} | >{$}c<{$} | >{$}c<{$} | @{}}
\hline
\text{Number}	    & \text{Rank} & \text{Number}     & \text{Paper} \\
\text{of citations} &		  & \text{of authors} & \text{description}
\\ \hline
c_1 & 1 & a_1 & p_1   \\
c_2 & 2 & a_2 & p_2   \\
\vdots & \vdots & \vdots & \vdots  \\
c_n & n & a_n & p_n
\\ \hline
	\end{tabular}
\noindent
   \begin{minipage}{0.65\textwidth}
\caption{
 \label{Table_4.2}
Citation list including the number of authors.
\newline \small
Here $a_i,\ i=1,\dots,n$ is the number of authors of the paper $p_i$.
}
   \end{minipage}
 \end{table}

	How we should proceed if there is at least one paper with at least two
authors? It is intuitively clear that in such a case the personal impact
(''fame'') of a particular author should be connected somehow with his/her
contributions in a multiple author paper (see the discussion on this item
in~\cite[page~4, case~3)]{Pendlebury-2010}. Generally we can distinguish the
following main cases.
    \begin{enumerate}
\item
	The authors do not supply any information about their personal
contributions in their joint paper or they write that these contributions
cannot be distinguished.
 \item
	It is explicitly said which parts of the work by who of the coauthors
are personally written.
 \item
	The authors present concrete information about their contributions in a
form of numbers.
    \end{enumerate}

	Evidently, there may be many other cases, \eg different parts of a work
realize some/all of the above three possibilities. As we do not want to overload
the presentation with too much details, we shall restrict our consideration to the
above cases.

	The most clear is case~3. Suppose we talk about paper $p_i$ of
table~\vref{Table_4.2} for some fixed $i$. Then to the $j$-th, $j=1,\dots,a_n$,
coauthor corresponds a number (weight) $w_j^{a_i}$ such that $0<w_j^{a_i}<1$,
$\sum_{j=1}^{a_i}w_j^{a_i}=1$ and the contribution of the $j$-th author is
exactly $w_i^{a_i}$.

	The complete lack of information about personal authors contributions
in case~1  leads to only one hypothesis for rigorous analysis, namely that all
coauthors have equal contribution in the work. This hypothesis, which we
assume, reduces case~1 to case~3 with $w_j^{a_i}=1/a_i$.

	Case 2 does not supply sufficient information for a rigorous analysis.
For example, a judgement of an author's contribution by the number of pages
he/she has written is not serious. Our intension is to reduce this case to
case~3 but there is not enough information to do this. So again, we shall assume
that $w_j^{a_i}=1/a_i$. However, regardless of the equalization of authors
contributions, the information given in case~2 may lead to some consequences
for our next considerations.

	We shall call the numbers $w_j^{a_i}$ \emph{personal authors weights}.
We assume that $w_j^{a_i}=1$ for $a_i=1$ to cover also the single-author case.

	The general approach to the fractionalizing and weighting the number of
publications and of the citations is outlined
in~\cite[pp.~22--23]{Rehn_and_Kronman-2008}.

	Let us now return to a citation list form from the viewpoint of the
contributions of the author to whom it belongs. Taking into account the above
discussion, we should add to the citation list a new column containing in its
$i$-row the personal author weight $w_i^{a}$ for the paper $p_i$. At this
point it becomes evident that not all of the fame for the paper $p_i$ having
$c_i$ citations belongs to the considered author if $a_i>1$, \ie for
$w_i^{a}<1$. Since the number $w_i^{a}$ is the only measure for the author's
particular contribution, we shall assume that from all $c_i$ citations of the
paper $p_i$ only the part $c_i^{a}:=w_i^{a}c_i$ belong to that author. We
shall call the numbers $c_i^{a}:=w_i^{a}c_i$ \emph{(author\ndash)reduced
number of citations} of the paper $p_i$. Its inclusion in a citation list leads
to the Table~\vref{Table_4.3} as a new form of citations lists.

\renewcommand{\arraystretch}{1.1}	

\begin{table}[ht!]


\vspace{1ex}
\centering
	\begin{tabular}
		{@{} | >{$}c<{$} | >{$}c<{$} | >{$}c<{$} | >{$}c<{$}
		     | >{$}c<{$} | >{$}c<{$} | @{}}
\hline
\text{Number}	    & \text{Rank} & \text{Number}     & \text{Author}
& \text{Reduced number} & \text{Paper}
\\
\text{of citations} &		  & \text{of authors} & \text{weight}
& \text{of citations}   & \text{description}
\\ \hline

c_1 & 1 & a_1 & w_1^{a} & c_1^{a}=w_1^{a}c_1 & p_1 \\
c_2 & 2 & a_2 & w_2^{a} & c_2^{a}=w_2^{a}c_2 & p_2 \\
\vdots & \vdots & \vdots & \vdots & \vdots & \vdots  \\
c_n & n & a_n & w_n^{a} & c_n^{a}=w_n^{a}c_n & p_n \\
\hline
	\end{tabular}
\noindent
   \begin{minipage}{0.80\textwidth}
\caption{
 \label{Table_4.3}
Citation list including data for author personal contributions.
\newline \small
Here $w_i^{a},\ i=1,\dots,n$, is the personal author weight for the paper
$p_i$.
}
   \end{minipage}                                          \end{table}

	Now the reduced citation numbers $c_i^{a}$ play the role of
the citation numbers $c_i$ at the beginning of this section, so we shall
rearrange table~\vref{Table_4.3} by their descending order and will introduce
the reduced rank that numbers the rows of the rearranged table. In this way we
obtain Table~\vref{Table_4.4} as a new form of a citation list.

\renewcommand{\arraystretch}{1.1}	

\begin{table}[ht!]


\vspace{1ex}
\centering
	\begin{tabular}
		{@{} | >{$}c<{$} | >{$}c<{$} | >{$}c<{$} | >{$}c<{$}
		     | >{$}c<{$} | >{$}c<{$} | >{$}c<{$} | @{}}
\hline
\text{Reduced number} & \text{Reduced} & \text{Number}
& \text{Rank} & \text{Number}     & \text{Author} & \text{Paper}
\\
\text{of citations}   & \text{rank}     & \text{of citations}
&		  & \text{of authors} & \text{weight} & \text{description}
\\ \hline
c_{k_1}^{a} = w_{k_1}^{a}c_{k_1} & 1 & c_{k_1}
& r_{k_1} & a_{k_1} & w_1^{a} & p_{k_1}
\\
c_{k_2}^{a} = w_{k_2}^{a}c_{k_2} & 2 & c_{k_2}
	& r_{k_2} & a_{k_2} & w_2^{a} & p_{k_2}
\\
\vdots & \vdots & \vdots & \vdots & \vdots & \vdots & \vdots
\\
c_{k_n}^{a} = w_{k_n}^{a}c_{k_n} & n & c_{k_n}
& r_{k_n} & a_{k_n} & w_n^{a} & p_{k_n}
\\
\hline
	\end{tabular}
\noindent
   \begin{minipage}{0.95\textwidth}
\caption{
 \label{Table_4.4}
Citation list arranged by descending order of the reduced number of citations.
\newline \small
Here $(r_1,\dots,r_n)$ and $(k_1,\dots,k_n)$ are permutations of $(1,\dots,n)$
and $c_{k_i}^{a} \ge c_{k_{i+1}}^{a}$, $i=1,\dots,n-1$.
}
   \end{minipage}
 \end{table}

	From table~\vref{Table_4.4} can be drown conclusions similar to the
ones at the beginning of this section, but now covering the multiple author
case.

	If $c_i=0$ for $i\ge n_0$ for some $n_0\in\{1,\dots,n\}$, then
$c_i^{a}=w_i^{a}\cdot 0=0$. For this reason the works with zero citations
sit at the bottom of table~\vref{Table_4.4} and their relative order from
table~\vref{Table_4.3} can be preserved.

\subsection{Citations in papers with multiple authors}
\label{Subsect4.2}

	The consideration of the number of authors of the citing papers leads to
other form of citation lists that reveals in a finer way the impact of the
author of the cited papers on (other) authors. The simple number of citations
of a work shows only how many times it has been used in other works. However,
it is not one and the same when a citing paper has one or more than one
authors. It is reasonable to suppose that all authors of a citing paper have
equal acquaintance with all references contained in it if it is not stated
explicitly something else in the paper. Assuming this hypothesis, we see that
the impact of a paper on a work citing it can be measured not only the number~1
(representing only the fact of citation) but more precisely by the number of
authors of the citing work each of which we suppose to know the cited paper and
have some benefit of it. Similarly, the number of authors of all papers citing
a given work can be taken as a measure of the influence of the cited work.~%
\footnote{~%
Some of the citing authors may coincide. %
}
    \begin{Rem}    \label{Rem4.1}
There are works whose number of authors may be classified as ''quite
large''. Examples of such papers can be found in the region of experimental
physics of elementary particles, where can be found papers with, say, 100--150
and more authors; for instance, in the work~\cite{2500_Authors_Paper} we see
more than~2500 authors. Usually as authors of such works are pointed whole
experimental collaborations. We do not want to speculate on how such works are
written and what is the particular contribution of their authors and so on.
However, it seems that the hypothesis of acquaintance of all authors with all
references breaks down for works with ''quite large'' number of authors.
    \end{Rem}

    \begin{Rem}    \label{Rem4.2}
It seems that as a ''normal'' upper limit on the number of authors of a research
paper or a book/monograph can be taken~7 or~4 respectively. With some reserve we
may replace these numbers by~9 and~6 respectively. Our opinion is that the
hypothesis  of equal acquaintance of all authors with all references is not true
for research articles or books whose number of authors is greater than~9 or~6
respectively. Similar (statistical) limits may be pointed and for other types
of publications such as review article or articles in encyclopedias. In any
way, if the number of authors of a work is greater then some ''reasonable''
number, which should depend on works types, then the mentioned hypothesis seems
not to be valid.
    \end{Rem}

    \begin{Rem}    \label{Rem4.3}
When the hypothesis of equal acquaintance of all authors of a work with all
references in it is not true and there is not other information concerning the
acquaintance of the authors with the references, we cannot make any conclusions
on the impact of a cited work (and its authors) on the authors of the citing
work based on the fact of citation. In such cases we shall consider the citing
work as written by only one author for the purposes of our analysis.
    \end{Rem}

	So, to any citing paper we assign a number, citing paper impact,
which is equal to the number of its authors that are acquainted with the cited
paper or to~1, if such an information is missing in the citing paper or cannot
be found by means of some reasonable hypotheses.~%
\footnote{~%
The standard case is to set the mentioned number equal to one which represents
only the fact of citation. We consider this situation quite rough.%
}
The sum of citing papers impact numbers for all (known) papers citing a work
will be called citation impact number of the cited work and will be denoted by
$c^i$. By adding these numbers to Table~\vref{Table_4.3} we obtain
Table~\vref{Table_4.5} as a new version of a citation list.

\renewcommand{\arraystretch}{1.1}	

\begin{table}[ht!]


\vspace{1ex}
\centering
	\begin{tabular}
		{@{} | >{$}c<{$} | >{$}c<{$} | >{$}c<{$} | >{$}c<{$}
	     	     | >{$}c<{$} | >{$}c<{$} | >{$}c<{$} | >{$}c<{$} | @{}}
\hline
\text{Number}	    & \text{Rank} &\text{Cit.\ impact}
& \text{Number}     & \text{Author}
	& \text{Reduced} & \text{Reduced im-} & \text{Paper}
\\
\text{of cit.} &		  & \text{number}
& \text{of authors} & \text{weight}
	& \text{cit. number}   & \text{pact number} & \text{descr.}
\\ \hline

c_1 & 1 & c_1^i & a_1 & w_1^{a} & c_1^{a}=w_1^{a}c_1 & I_1=w_1^ac_1^i & p_1 \\
c_2 & 1 & c_2^i & a_2 & w_2^{a} & c_2^{a}=w_2^{a}c_2 & I_2=w_2^ac_2^i & p_2 \\
\vdots & \vdots & \vdots & \vdots  & \vdots & \vdots & \vdots & \vdots  \\
c_n & n & c_n^i & a_n & w_n^{a} & c_n^{a}=w_n^{a}c_n & I_n=w_n^ac_n^i & p_n \\
\hline
	\end{tabular}
\noindent
   \begin{minipage}{0.80\textwidth}
\caption{
 \label{Table_4.5}
Citation list including data for citation impact numbers.
\newline \small
The reduced impact numbers $I_j=w_n^ac_j^i$, $j=1,\dots,n$, take into account
the author contribution weights as well as the citation impact numbers.
}
   \end{minipage}                                          \end{table}

	If the reduced impact citation numbers $I_j=w_n^ac_j^i$, $j=1,\dots,n$,
can be introduced, then we can rearrange Table~\vref{Table_4.5} by
their descending order and call the number of a row of the so-obtained table
the \emph{reduced impact citation rank} of the paper sitting in it. In this way
we obtain the Table~\vref{Table_4.6} below as new modified version of
Tables~\vref{Table_4.5} and~\vref{Table_4.4} .

\renewcommand{\arraystretch}{1.1}	

\begin{table}[ht!]


\vspace{1ex}
\centering
	\begin{tabular}
		{@{} | >{$}c<{$} | >{$}c<{$} | >{$}c<{$} | >{$}c<{$}
		     | >{$}c<{$} | >{$}c<{$} | >{$}c<{$} | >{$}c<{$}
		     | >{$}c<{$} | @{}}
\hline
 \text{Reduced} & \text{Red.} & \text{Reduced}
	& \text{Rank \&} & \text{Number}
		& \text{Cit.} & \text{Number} & \text{Personal} & \text{Paper}
\\
\text{impact cit.} & \text{cit.} & \text{number}
	& \text{Reduced} & \text{of}
		& \text{impact} & \text{of} &\text{author} & \text{desc-}
\\
\text{number} & \text{rank} &\text{of citations}
	& \text{rank}  & \text{cit.}
		& \text{number} & \text{authors} & \text{weight} & \text{rition}
\\ \hline
I_{m_1}=w_{m_1}^a c_{m_1}^i & 1
& c_{m_1}^{a} = w_{m_1}^{a}c_{m_1} & r_{m_1} \text{ \& } r_{m_1}^r & c_{m_1}
& c_{m_1}^i & a_{m_1} & w_1^{a} & p_{m_1}
\\
I_{m_2}=w_{m_2}^a c_{m_2}^i & 2
& c_{m_2}^{a} = w_{m_2}^{a}c_{m_2} & r_{m_2} \text{ \& } r_{m_2}^r & c_{m_2}
& c_{m_2}^i & a_{m_2} & w_2^{a} & p_{m_2}
\\
\vdots & \vdots & \vdots & \vdots & \vdots & \vdots & \vdots  & \vdots & \vdots
\\
I_{m_n}=w_{m_n}^a c_{m_n}^i & n
& c_{m_n}^{a} = w_{m_n}^{a}c_{m_n} & r_{m_n} \text{ \& } r_{m_n}^r & c_{m_n}
& c_{m_n}^i & a_{m_n} & w_n^{a} & p_{m_n}
\\
\hline
	\end{tabular}
\noindent
   \begin{minipage}{0.95\textwidth}
\caption{
 \label{Table_4.6}
Citation list arranged by descending order of citation impact numbers.
\newline \small
Here $(m_1,\dots,m_n)$, $(r_1^r,\dots,r_n^r)$ and $(r_1,\dots,r_n)$ are
permutations of $(1,\dots,n)$ and by definition $I_{i} \ge I_{i+1}$ for
$i=1,\dots,n-1$.
}
   \end{minipage}
 \end{table}

	In conclusion, we have three major forms of any citation list which are
given via the tables~\vref{Table_4.2}, \vref{Table_4.4} and~\vref{Table_4.6}
which are suitable for farther analysis of the data in them.


\section {Bibliometric indices (metrics)}
\label{Sect5}

	The bibliometric indications~\cite{Rehn_et_al-2007} are a known tool
for measuring authors impact. Starting from 2005 there ware introduced many
new (bibliometric) indices, called also metrics, whose purpose is to measure the
influence of an author on the ground of citations of his/her works. These
indices can be described as bibliometric and their connection with the
scientific impact of an author is indirect~%
\footnote{~%
It is based on statistical data analysis~\cite{h-index_and_9_its_variants,
h-index_and_37_its_variants, h-index_for_147_chemistry_research_groups}.%
}
as it cannot be revealed without knowing the content of the cited and citing
papers. However, the usage of these indices has brought significant advance in
this area compared to the previous analysis based, for instance, on author's
total number of published works and their total number of citations. For
example, in~\cite{Jensen_et_al-2008} are provided arguments that ''the number
of citations or the mean number of citations per paper are definitely not good
predictors of promotion''.

	This section aims to list a few bibliometric indices and to present
some analysis on their ground. It is not our goal here to present a
''complete'' list of (all) bibliometric indices introduced until now as well
as to point to their ''good'' and ''bad'' sides, which are known and
already described (see, e.g.,~\cite{Alonso_et_al-2009},
http://sci2s.ugr.es/hindex/biblio.php and http://sci2s.ugr.es/hindex/).

\subsection{The Hirsch index}
\label{Subsect5.1}

	All of the new game started  with 2005 paper of
J.~Hirsch~\cite{Hirsch-2005} in which he defined the $h$\ndash index, called
nowadays the Hirsch index, as follows.
    \begin{quote}
A scientists has index $h$ if $h$ of his/her $N_p$ published papers have at
least $h$ citations each and the other $(N_p-h)$ papers have no more than $h$
citations each.
    \end{quote}
(This is not the Hirsch original definition, but the one of  September 2006
e\ndash print.) In terms of Table~\vref{Table_4.1}, we have
    \begin{equation}    \label{5.1}
c_h \ge h \ge c_{h+1}
    \end{equation}
\ie $h$ is the maximal rank  such that the corresponding to it paper has no
less than $h$ citations and the papers with greater ranks have maximum $h$
citations. The author of the present paper failed to find in the available to
him literature arguments why the Hirsch index was defined exactly in this way.
It contains only discussions of the pros and cons of the Hirsch index (see,
for instance, the discussion of the Hirsch index
in~\cite[Sect.~1]{Costas_and_Bordons-2007} and in~\cite{Alonso_et_al-2009,
Bornmann_and_Daniel-2007, Kelly-Jennions-2006}). Of course, the pros are a
posteriori arguments of the definition but they do not answer the question why
it works (''well'') in some cases.~%
\footnote{~%
 The Hirsch index is applicable also for groups of scientist united by a
journal, country, institute/university \etc For instance, in the site
http://www.scimagojr.com/ it is calculated for the journals and countries
covered by the Scopus database with URL http://www.scopus.com.%
}
The Hirsch index received a lot of attention and found many applications as it
combines in a single number quality, productivity and impact of an author. In
general it correlates with other bibliometric
indices~\cite{Kulasegarah_and_Fenton-2010}.

	By our opinion, one of the ideas behind the $h$\ndash index is the
selection of some of the ''top cited'' papers of an author and to take their
number as a measure of his/her publications impact which is confirmed
a posteriori by the results in~\cite{Schreiber-2010}.~%
\footnote{~%
Alternately, one can take as a measure, for instance, the number of papers
with at least $N$ citations or the number of citations of all papers with rank
greater or equal to $M$ for some integers $N$ and $M$. However, the numbers $N$
and $M$ are arbitrary to a great extend irrespectively of are they constant or
not with respect to all authors. Example of such a measure is the ''Einstein
index'' (see
$http://www.science20.com/hammock\_physicist/who\_todays\_einstein\_exercise\_ranking\_scientists-75928$)
characterized by $M=3$.%
}
From this point of view the Hirsch index has two significant advantages: (i)~it
adapts to any particular author, hence being author\ndash dependent and (ii)~it
naturally defines the top cited papers as ones whose number of citations is no
less that it.

	There can be defined many indices that have the same properties as the
Hirsch index. For example, we can define an $f$\ndash modified Hirsch index
$h^f$ for some function $f \colon \field[R]^+\to \{1,\dots,n\}$ (in the
notation of Table~\vref{Table_4.1}) via
    \begin{equation}    \label{5.2}
c_{h^f} \ge f(h^f) \ge c_{h^f+1}
    \end{equation}
for particular choices of $f$; for example, $f(h^f)=h^f+1$ and $f(h^f)=h^f-2$
lead to different indices~%
\footnote{~%
The Hirsch index is selected by $f(h^f)=h^f$.%
}
whose usefulness can be determined only by making particular calculations for
particular authors. Without going into details we shall say that the results
strongly depend on $f$ and generally are not ''stable'' with respect to the
choice of $f$. Similarly, if we take a function
$g \colon\{1,\dots,n\} \to \field[R]^+ $, which may be the one inverse to $f$
if it exists, then we can rewrite~\eref{5.2} as
    \begin{equation}    \label{5.3}
c_{g(h_g)} \ge h_g \ge c_{g(h_g+1)}
    \end{equation}
which introduces other modification $h_g$ of the Hirsch index. The particular
choice $g(h_g)=10h$ reproduces the $w$\ndash index~\cite{Wu_Q-2010}. Similarly
can be obtained the $k$\ndash\ and $w$\ndash indices as defined
in~\cite{Anania-Caruso-2012}.

\subsection{Modifications of the Hirsch index}
\label{Subsect5.2}

	The Hirsch index does not reflect many important data contained in a
citation list. This has lead to the introduction of a lot of its variants each
of which tries to take into account some features which the original Hirsch index
misses to reflect. An excellent review on the Hirsch index and many its variants
can be found in~\cite{Alonso_et_al-2009}. A list of~37 versions of the Hirsch
index is contained in~\cite[Table~1 on page~349]{h-index_and_37_its_variants}
(see also~\cite{h-index_and_9_its_variants}) which paper contains also a quit
complete list of relevant references. In~\cite{Schreiber-2010} are analyzed and
calculated~20 versions of the Hirsch index. Below we shall pay attention to some
of the modifications of the Hirsch index that are closer to the aims of this
work.

\subsubsection{Multiple authorship}
\label{SubSubsect5.2.1}

	The Hirsch index $h$ is insensitive to how many authors have the papers
in Table~\vref{Table_4.1}. But this index aims to represent the contribution of
a particular author whose citation list is considered. So, if some or all of
the first $h$  papers in Table~\vref{Table_4.1} have more than one author, then
it is evident that in the $h$\ndash index is incorporated also the work of
authors different form the one whose list of citations is investigated. The
correction of this unfairness with respect to the other authors (whose work is
assigned to other person(s)) leads to a class of indices that reflect the
number of authors of the cited papers. For definition and analysis of such
indices are suitable citation lists in a form given by Table~\vref{Table_4.2}.

	The $h_m$ index introduced by Schreiber~\cite{Schreiber-2008} is
defined via equation~\eref{5.3} with the choice
$g=r_{\text{eff}}^{-1} \colon \field[R]^+ \to \{1,\dots,n\}$
for
    \begin{equation}    \label{5.4}
r_{\text{eff}}^{-1} \colon r \mapsto r_{\text{eff}}^{-1}(r)
=
\sum_{i=1}^{r}\frac{1}{a_i}
    \end{equation}
where $r\in\{1,\dots,n\}$, $r_{\text{eff}}^{-1}$ is treated as an effective
rank of the paper $p_r$ and we use the notation of Table~\vref{Table_4.2}. We
should mention that here is used the hypothesis of equal contribution of all
authors of a multiple author paper. In~\cite{Schreiber-2009} the
$h_m$\ndash index is calculated for~26 particular cases, which shows strong
correlation with the $h$\ndash index but the arrangement of the
authors according to the both indices is generally quite different.

	In the more general case, when personal authors weights are known (see
Table~\vref{Table_4.3}), the function $g$ in~\eref{5.3} should be chosen as
$g=r_w^{-1}$ with
    \begin{equation}    \label{5.5}
r_w^{-1} \colon r \mapsto r_w^{-1}(r) = \sum_{i=1}^{r} w_i^a
    \end{equation}
which reduces to~\eref{5.4} for $w_i^a=1/a_i$ and leads to the
author\ndash weighted $h_w^a$\ndash index. Thus we have
    \begin{align}    \label{5.6}
c_{r_{\text{eff}}^{-1} (h_m) } & \ge h_m \ge c_{r_{\text{eff}}^{-1} (h_m+1) }
\\    \label{5.7}
c_{r_w^{-1} (h_w^a) } & \ge h_w^a \ge c_{r_w^{-1} (h_w^a+1) }
    \end{align}
The values $w_i\equiv 1$ reduce $h^a_w$ to the original Hirsch $h$\ndash index.

	The $h_I$\ndash index~\cite{Batista_et_al-2006} corrects the $h$\ndash
index by dividing it by the mean number of authors of papers selected by
the $h$\ndash index,
    \begin{equation}    \label{5.8}
h_I = h/\bar{a}, \quad \bar{a} := \bigl(\sum_{i=1}^{h}a_i\bigr)/h
    \end{equation}
in the notation of Table~\vref{Table_4.2}.

	In the Publish or Perish program user manual~%
\footnote{~%
See http://www.harzing.com/pophelp/metrics.htm.%
}
is defined the normalized Hirsch index $h_{I,\text{norm}}$ (Individual normalized
Hirsch index) which is defined similarly to the Hirsch index with the difference
that now is used Table~\vref{Table_4.4} and it is supposed that $w_i^a=1/a_i$,
\ie (cf.~\eref{5.1})
    \begin{equation}    \label{5.9}
c_{h_{I,\text{norm}}}^a  \ge h_{I,\text{norm}} \ge c_{h_{I,\text{norm}}+1}^a .
    \end{equation}
In words, the papers are ordered by the descending order of the citations
divided by the corresponding number of authors and then the (normalized) Hirsch
index is calculated. The author of these lines shares the opinion that the
$h_{I,\text{norm}}$\ndash index reflects the author achievements
considerably better than the original Hirsch index and the $h_m$\ndash index.

	The below introduced by~\eref{5.16} AWCRpA\ndash index also takes care
of the number of authors of the cited papers.

\subsubsection{Taking into account missed citations}
\label{SubSubsect5.2.2}

	The only information about the number of citations contained in the
Hirsch index $h$ is that their total number is no less than $h^2$
(see~\eref{5.1}). It is clear that the more citations a paper has, the more
weight it should be given and \vv~%
\footnote{~%
Unfortunately the Hirsch and Hirsch-like indices completely lost the low cited
papers with non\ndash zero  citations, \eg the ones with less than $h$
citations in a case of the $h$\ndash index.%
}
The $g$\ndash index~\cite{Egghe-2006} and the $e$\ndash index~\cite{Zhang-2009}
aim to correct this situation with the Hirsch index.

	The $g$-index of an author with citations list like
Table~\vref{Table_4.1} is the unique largest number $g$ such that the total
number of citations of the first $g$ papers is greater than or equal to $g^2$.
Its aim is to give more weight to papers with more citations and thus improving
the $h$\ndash index.

	The $e$-index also gives more attention to highly cited works and also
helps to make difference between authors with similar Hirsch indices but
different citations numbers. Using again the notation of
Table~\vref{Table_4.1}, we have
    \begin{equation}    \label{5.10}
e = \sqrt{\sum_{i=1}^{h}(c_i-h) }
  = \sqrt{\sum_{i=1}^{h}c_i - h^2 }
    \end{equation}
where $h$ is the Hirsch index of the author. The $e$\ndash index is
complementary to the  $h$\ndash index as it gives/measures some of the
citations missed by the Hirsch index.

	Similar aims persuade also:~%
\footnote{~%
See~\cite[table~2 on page~829]{Karpagam_et_al-2011} and the references given
therein.%
}
the $h^2$\ndash index,
the $A$\ndash index ($=\frac{1}{h}\sum_{i=1}^{h}c_i$),
the $R$\ndash index (=$\sqrt{Ah}$), the $h_{\text{w}}$\ndash index,
and the $hg$\ndash index ($=\sqrt{gh}$).

	The citations outside the $h$\ndash index core are taken into account
also in the indices introduced in the following sub\ndash subsections.

\subsubsection{The time dependence}
\label{SubSubsect5.2.3}

	Until now we have not touched the problem for the dependence of the
citations on the time. The simples way to fill this gap is the introduction of
the age of the cited papers.

	Suppose we have a citation list in a form of Table~\vref{Table_4.1} and
$t_i$ is the age of the paper $p_i$, $i=1,\dots,n$, counting from its first
publication. Then the $AR$\ndash index is
    \begin{equation}    \label{5.11}
AR = \sqrt{\sum_{i=1}^{h}c_i/t_i}
    \end{equation}
with $h$ being the Hirsch index of the considered author. The $AR$\ndash index
may decrees with time.

	The contemporary $h$\ndash index
$h^c$~\cite[Sect.~2]{Sidiropoulos_et_al-2006} is defined similarly but instead
of the number $c_i$ of citations of the paper $p_i$ is used the score
    \begin{equation}    \label{5.12}
S^c(i) = \gamma c_i/(1+t_i)^\delta
    \end{equation}
where $\gamma$ and $\delta$ are constants and $t_i$ is the paper age in years
(counted from its publication); often is taken $\gamma=4$ and $\delta=1$. An
author has index $h^c$ if $h^c$ of his/her papers have a score not less than
$h^c$ and the remaining ones have a score not greater than $h^c$. In
particular, if we arrange a citation list by descending values of $S^c(i)$, then
(cf.~\eref{5.1})
    \begin{equation}    \label{5.13}
S^c(h^c) \ge h^c \ge S^c(h^c+1).
    \end{equation}
If the score~\eref{5.12} is modified as
$S^c(i) = \gamma \sum_{t\in c_i} 1/(1+t)^\delta$
we obtain the trend $h$\ndash index~\cite[Sect.~2]{Sidiropoulos_et_al-2006}.

	In the program Publish or Perish are introduced three other indices
that depend on the age of the cited work.~%
\footnote{~%
See http://www.harzing.com/pop.htm and
http://www.harzing.com/pophelp/metrics.htm.%
}
The age\ndash weighted citation
rate is
    \begin{equation}    \label{5.14}
AWCR = \sum_{i=1}^{n} c_i/t_i
    \end{equation}
where $c_i$ and $t_i$ are the citations and the age of the $i$-th paper and
the sum is over all published papers, and the age\ndash weighted index is
    \begin{equation}    \label{5.15}
AW = \sqrt{AWCR} = \sqrt{\sum_{i=1}^{n} c_i/t_i}.
    \end{equation}
Note that~\eref{5.15} differs from~\eref{5.11} by the inclusion of citations
outside of the $h$\ndash core. If the paper $p_i$ has $a_i$ authors, then the
per\ndash author modification of~\eref{5.14} is
    \begin{equation}    \label{5.16}
AWCRpA = \sum_{i=1}^{n} c_i/(t_i a_i) .
    \end{equation}

\subsection{Comments}
\label{Subsect5.3}

	As we have seen, there were introduced quite a number of bibliometric
indices. Their properties are well known and discussed at length in the cited
references and the ones given in them. The general opinion is that different
indices represent different measures of author's published works and in many
cases are complimentary to each other. This points to the complexity of the
problem of giving an evaluation of authors impact by using citation
lists.

\subsection{Generation of new indices}
\label{Subsect5.4}

	In subsection~\ref{Subsect5.1} we  pointed that to functions
\[
f \colon \field[R]^+ \to \{1,\dots,n\}
\qquad
g \colon \{1,\dots,n\} \to \field[R]^+
\]
(we use the notation of Table~\ref{Table_4.1}) there correspond respectively
indices $h^f$ and $h_g$ with values in $\{1,\dots,n\}$ such that
    \begin{subequations}    \label{5.17}
    \begin{align}    \label{5.17a}
 c_{h^f} & \ge f(h^f) \ge c_{h^f+1}
\\    \label{5.17b}
 c_{g(h_g)} & \ge h_g \ge c_{g(h_g)+1} .
    \end{align}
    \end{subequations}
Here we implicitly supposed that the functions $f$ and $g$, which may be inverse
to each other, are such that $h^f$ and $h_g$ exist and are unique which puts
some restrictions on these functions. These are more or less trivial versions
of the Hirsch index (cf.~\eref{5.1}) regardless that their particular
properties and interpretation may be quite different depending on the
particular choices of $f$ and $g$.

	When Hirsch-like indices are utilized, only part of the author's papers
are taken into account. An important moment is that the number of these papers
is author\ndash dependent. Often, as in the case of the Hirsch index, this
selection is done by the rank (sequential number) of the papers in a citation
list in which the papers are arranged by descending number of citations
(possibly normalized by some factors/weights). However, there are infinite
number of ways to make similar selections on the base of other principles.~%
\footnote{~%
Take, for instance, a citation list of a form of Table~\ref{Table_4.4}. For
$w_i\equiv 1$ it is a base for defining the $h$\ndash index and for $w_i=1/a_i$
is a base for the introduction of the $h_{I,\text{norm}}$\ndash index.%
}

	Define the (arithmetic) mean of the non-vanishing reduced numbers of
citation by (we use the notation of Table~\vref{Table_4.4})
    \begin{equation}    \label{5.18}
\bar{c}^a
=
\frac{\sum_{i=1}^{n}c_i^a}
     { \sum_{i\in\{1,\dots,n\}, \ c_i\not=0 }^{} 1} .
    \end{equation}
Now we can define a new index, say $\bar{h}^a$, via (cf.~\eref{5.1})
    \begin{equation}    \label{5.19}
\bar{h}^a = \max_{r\in\{1,\dots,n\}}\{r: c_r^a \ge \bar{c}^a\} ,
    \end{equation}
\ie $\bar{h}^a$ selects the papers with at least $\bar{c}^a$ citations and it
equals to the maximal reduced rank between papers with this property.
Evidently, we can replace $\bar{c}^a$ with other mean values, \eg with the
geometric mean value of all papers with non\ndash vanishing citations, and will
obtain in this way a new index like $\bar{h}^a$ above. One can even use the
mean square deviation
\[
\delta = \sqrt{ \sum_{c_i^a\ge \bar{c}^a}^{} (c_i^a-\bar{c}^a) }
\]
to define highly cited papers by $c_i^a\ge\bar{c}^a+\delta$ and use this
inequality in the r.h.s.\ of~\eref{5.19} to define a new index.

	Another way for generation of new Hirsch-like indices is to redefine the
existing ones, usually based on tables~\ref{Table_4.1}, \ref{Table_4.2}
and~\ref{Table_4.3}, by indices based on tables~\ref{Table_4.4},
\ref{Table_4.5} and~\ref{Table_4.6}. We do not want to go into details of
this process as it is quite clear and evident and the real problem is how
useful the new\ndash obtained indices will be, which can be solved only by
making particular calculations for particular persons. In any case, our
opinion is that indices based on tables~\vref{Table_4.4} and~\vref{Table_4.6}
should be better than the original ones.

	From theoretical point of view it can be invented an infinite umber of
''indices'' that will reflect different aspects of a citation list. The
discussed in the literature bibliometric indices confirm this opinion.

\subsection{Which is the best index?}
\label{Subsect5.5}

	An analysis of some bibliometric
indices~\cite{h-index_and_37_its_variants,
h-index_and_9_its_variants, Costas_and_Bordons-2007} reveals that any one of
them has its pros and cons and is useful in some cases and gives unsatisfactory
consequences in other ones. All this points that there cannot be pointed the
''best index'' unless there are well defined criterion(s) what it must
satisfy, what is expected from it and what is the area of its application.
For example, if we are interested simply of the impact of a paper, then, e.g.,
the $h$\ndash index is better then the $h_m$ and $H_{I,\text{norm}}$ indices,
but if we aim to evaluate the author personal (individual) impact, then the
$h_m$ and $H_{I,\text{norm}}$ indices are more adequate than the Hirsch index.
Similarly, we have an intuitive understanding of ''highly cited'' papers of an
author but without a rigorous definition of this concept we cannot do much. The
same is the situation with the ''low cited'' papers with non-vanishing number
of citations. Besides, there is a problem why some or all of the ''low cited''
papers are excluded from the scope of some of the bibliometric indices like the
Hirsch index and most of the Hirsch-like ones.

	The above point to the complexity of the problem of citation analysis
and author evaluation/impact based on it. As we said, we share the opinion
that the known approaches to it reveal only some its aspects and no one of them
gives a ''complete'' answer. Besides, we agree in general with
Hirsch~\cite[p.~4]{Hirsch-2005} that ''a single number can never give more than a
rough approximation to an individual's multifaced profile'', but this concerns a
more general problem than the one investigated in this work.

\subsection{What to do next?}
\label{Subsect5.6}

	Tens of bibliometric indices are in current
usage~\cite{h-index_and_37_its_variants}. The process of invention and testing
of new indices can be continued with a hope that the ''best'' index will be
found.

	The final goal is to be found quantitative measures for evaluation and
comparison of authors and their impact. At the moment we consider the case when
the information for realization of this aim are the citation lists of the
authors. In this respect we notice that citation impact is strongly influenced
by the following factors~\cite[p.~61]{Glanzel-2003}:
	(i)~the subject matter and within the subject, the ''level of
abstraction'', (ii)~the paper's age, (iii)~the paper's ''social status''
(through the author(s) and the journal), (iv)~the document type and (v)~the
observation period.
	All of them have to be taken into account when evaluating the
scientific impact of a scientists.

	There are two global characteristics of a citation list like the one
presented by Table~\ref{Table_4.1} that are often used: the total number $n$ of
published papers and the total number
    \begin{equation}    \label{5.20}
c:=\sum_{i=1}^{n} c_n .
    \end{equation}
 of their citations. To them can be added the author coefficient of citation
performance
    \begin{equation}    \label{5.21}
E = \bigl( \sum_{c_i\not=0} 1 \bigr) / n
  = \bigl( \sum_{c_i\not=0} 1 \bigr) / \bigl( \sum_{c_i} 1 \bigr)
    \end{equation}
which is the ratio of the number of papers with non-vanishing citations and the
number of all papers. From these numbers can be made qualitative conclusions
concerning authors like: the greater $n$, the more productive/active an author
is and the greater $c$, the more is his/her impact on (other) authors. Of
course, the coefficient of performance~\eref{5.21} is a rigorous measure but it
concerns only a single author and cannot be used to measure the authors impact
on other authors; it only measures now much of his/her works have non\ndash
vanishing usage by (other) authors.

	The total number of citations $c$ shows in how  many papers the author's
works have been mentioned/used. But, since we aim to make conclusions
concerning only the author, not his/her co\ndash authors, if any, this number
in the general case does not give adequate measure of the author without
counting the number of authors of each paper. For the purpose the index of
citations
    \begin{equation}    \label{5.22}
c_I = \sum_{i=1}^{n} c_i w_i
    \end{equation}
is considerably better characteristic. Here we have used the notation of
Table~\vref{Table_4.3}; recall, $w_i=1/a_i$ if all authors are suppose to have
equal contributions in the paper $p_i$. The $c_I$\ndash index shows how
many papers have been influenced by the author's personal contribution in
his/her published works as a whole.

	Notice, the weights $w_i$ can also be used to make the
effectiveness~\eref{5.21} more accurate, \viz by making it to represent the
author individual contributions:
    \begin{equation}    \label{5.21-1}
E^I = \bigl(\sum_{c_i\not=0}w_i\bigr) / \bigl(\sum_{i=1}^{n}w_i\bigr)
= n_{\not=0}^I / n^I,
    \end{equation}
where the numbers
    \begin{align}    \label{5.21-2}
n^I &:= \sum_{i=1}^{n}w_i
\\    \label{5.21-3}
n_{\not=0}^I &:=  \sum_{i\in\{1,\dots,n\}, \ c_i\not=0}w_i
    \end{align}
can be interpreted as respectively effective (individual) number of author's
published works and individual number of papers with non\ndash zero number of
citations.

	The index~\eref{5.22} can be called individual-to-works impact index.
Similarly, we can introduce individual\ndash to\ndash author index $c_I^A$ which
shows how many authors have been influenced by the publish papers of an author.
For the purpose we notice that in $c_I$ every citing paper is counted exactly
one time, which represents the simple fact of citation. Instead of this we can
count the number of authors of each citing paper that are acquainted with the
cited work, if these numbers are known. So, the analogue of the total citation
number~\eref{5.22} now is
    \begin{equation}    \label{5.23}
c^i := \sum_{i=1}^{n} c_j^i
    \end{equation}
where $c_j^i$, $j=1,\dots,n$, is the sum of all authors of all papers citing the
paper $p_j$ and which authors are acquainted with the cited paper $p_j$.~%
\footnote{~%
Recall, if for some citing paper cannot be determined the number of authors
acquainted with the cited paper, then we set this number equal to one, which
represents only the fact of citation.%
}
Besides, to take into account the contribution of the author whose citation
list is considered, we have to introduce in~\eref{5.23} also his/her weight
$w_j$ in the creation of the paper $p_j$. Hence the number
    \begin{equation}    \label{5.24}
c_I^i : = \sum_{j=1}^{n}  c_j^i w_i = \sum_{j=1}^{n} I_j
    \end{equation}
is a global measure of the individual impact of an author on (other) authors
citing his/her papers.

	The global characteristics like~\eref{5.20}--\eref{5.24} miss a lot of
local information and in this respect some of the existing or new bibliometric
indices may provide essential complimentary information. However, here are
needed particular calculation which is not a purpose of this work.


\section {To what should be paid attention?}
\label{Sect6}

	Until now we have looked on the citations from pure bibliometrics point
of view. However our aim at this stage is the usage of citation analysis for
making conclusions for the scientific impact/achivements of a scientist without
going into the scientific content of the cited and citing papers. In this
respect there are important arguments that are not purely bibliometric.

\subsection{Self-citations}
\label{Subsect6.1}

	When an author cites a paper and he/she is between the authors of this
paper, we say that this is a self\ndash citation for this author. Often this
definition is broaden by saying that a citation is a self-citation if the
intersection of the authors of the cited and citing works is not empty.

	As we said earlier in Subsection~\ref{Subsect2.5}, the self\ndash
citations may be included in the citation lists and treated on equal footing
with the rest of author's citations when the self-\ndash citations are made for
pure scientific reasons, in particular when an author uses his/her earlier
works in subsequent own publications.

	One of the problems with the
self-citations~\cite[Sect.~4.2]{Glanzel-2003} is that they can easily be
manipulate and, if this is the case, this artificially brings more citations.
So, in this context, the problem is more a moral than a scientific one and our
believe is that most of the authors are fair in this respect, do not cite their
own papers without need and cite them on equal footing with other works.

	Good reasons why the self-citation should not be count are given at the
beginning of~\cite[Sect.~V]{Schreiber-2007} and we agree with them.
Analysis of the self\ndash citations and their influence on some
bibliometric indices can be found in many papers like~\cite{Schreiber-2007,
Costas_Leeuwen_Bordons-2010, Bartneck_Kokkelmans-2011,
Gianoli_Molina-Montenegro-2009}.
	It seems that the general opinions are that the self-citation should be
excluded when evaluating the scientific impact of an author. The main reason
for this is that it is more important the influence of author's works on other
scientists and their papers than on the author himself/herself. Other
important fact confirming the exclusion of self\ndash citations is that they
can relatively easy be manipulated in favour of one or other author; as pointier
in~\cite{Schreiber-2007a} ''not only the author's own self-citations have a
substantial effect in reducing the Hirsch index appreciably, but also the
self-citations of the co-authors are usually quite significant and reduce the
Hirsch index further''.

	In this respect it seems reasonable to be introduced an upper limit
on the number of self\ndash citation which should be regarded as
natural/ordinary citation. This limit is reasonable to be connected with the
total number of author's publications and possibly with the number of their
citations. If the self\ndash citations exceed this limit, the self\ndash
citation should be reduced to it or neglected at all.

	In general, the approaches to self-citations
are~\cite[p.~21]{Rehn_and_Kronman-2008}: excluding them, noting them and
trying to give a suitable their interpretation and ignoring their effect by
assuming their even distribution.

	An example of a bibliometric index that takes into account
self-citations is the $V$\ndash index~\cite{Ferrara-Romero-2012}, which is the
$h$\ndash index multiplied by the square root of one minus the ratio of the
number of self-citations and the total number of citations.

\subsection{The number of authors}
\label{Subsect6.2}

	Since we wont to make conclusions about a particular author whose
citation list is analyzed, the number of authors of each his/her paper should be
counted and taken into account. More precisely, his/her contribution/weight in
any cited paper in his/her citation list should be presented and used for the
citation analysis. If such an information is missing, we assume that his/her
contribution in a paper is one divided by the number of all paper's authors.

	Another thing is the number of authors of citing works. If it is
insignificant for some problem, then such a citation simply adds the number one to
the number of citations to the cited paper. But if this number is important,
then instead of counting this paper once, we should consider replacing this
weight~1 by the number of its authors or, more precisely, the number of its
authors that are acquainted (and using) the cited paper, if the last number can
be determined.

\subsection{Highly/low cited papers}
\label{Subsect6.3}

	It is generally accepted that the more/less citations a paper has, the
more/less weight to its impact should be assigned. However, in the general case
there is no a definition or a criterion of highly/low cited paper except that
this seems to be author\ndash dependent concept and the first/last paper in
Table~\ref{Table_4.1} is regarded as heighly/low cited one (it is possible
that the both may coincide). A discussion of this problem with some
proposals for its solution is given in~\cite[pp.~71--72]{Glanzel-2003}.

	When we are dealing with Hirsch\ndash like indices, then the highly
cited papers are the ones in their cores, while the ones outside the cores are
low\ndash cited. and are not taken into account at all. Besides, in some cases,
the more citations has a paper in the core, the more weight it has in the
corresponding indices.

\subsection{Different editions/versions of a published work}
\label{Subsect6.4}

	From bibliometric point of view any separate publication of a work is a
different published paper no matter if there is a difference between the
publications. But there is also a different point of view.
	For example, two identical editions of a book are different
publications but form scientific position the second edition simply supplies
more copies of the book as it does not contain different content. Similarly, a
lot of works are published in journals and then appear in an identical form in
collections of papers. Besides, there are books/monographes translated in
different languages which, excluding some (introductory) remarks and comments by
the  publishers/translators, are identical by their content. Evidently these are
different publications that are copies of one and the same work form scientific
point of view. Moreover, a work can have essentially different publications
like preprints, electronic prints, conference reports and journal versions
that have identical content and differ possibly only in the presentation of
the material (\eg differently numbered equations and permutations of parts of
the text).

	So, we face a problem: in a citation list may be presented different
publications which are identical from scientific view point. Our opinion
is that such publications should be  identified under a single work title.

	Unfortunately there is not a strict criterion when two different
publications should be considered identical. For instance, adding essentially
new results to a previously published work can be considered as resulting into a
new work, but a renumbering/pemutating the sections and/or equations in it
does not change it for the Science.

\subsection{Different types of publications}
\label{Subsect6.5}

	In Section~\ref{Sect2} we talked about reasons for citation in
different kinds of publications. Now we can look on this material from two more
viewpoints: are there reasons for assigning different weights for citation in or
citing from different types of publications?  Here and below by publication
types we shall understand such as: monograph/book,  textbook, booklet, originals
research article, review paper, collection of original  or already published
papers, handbook, encyclopedia, simple list of papers on some subject(s), and
so on.

	To begin with, it must be mentioned that the publication type
(partially) reflects the paper content; \eg it is quite rare a research paper
to appear in an encyclopedia, but a review work may be published in a
textbook, handbook or encyclopedia.

	Consider first citing papers of different types. Are there reasons to
assign different weights to them without knowing their particular content? In
Section~\ref{Sect2} we paid attention to some reasons for citation in some
publication types. These reasons are generally type\ndash dependent and a
priority of some of them can be given only via additional hypothesis. Consider
for instance a citation in an original research work. If one wants simply to
cite in it a particular result, \eg an equation, then a random choice of a paper
containing it is sufficient. However, if the author wants not only to mention
work containing the result, but also to pay a tribute to author(s) that have
firs found it, then a priority will be given to papers that historically
mentioned the result for the first time. So, our opinion is that without
knowing the content of a citing paper we cannot assign in an abstract way
different weights to the cited papers in a sense that when a paper is cited in
different types of publications we cannot assign different weights to such a
citation without an additional information.

	Consider now cited papers of different types. It is a generally accepted
opinion, \eg in annual reports and personal CVs, that books/monographes weight
more then other publication types, a chapter in a book is heavier than a research
article and so on. But what are the reasons for such a rating? Nowadays a new
result normally appears first as a research (journal) article, (electronic)
preprint, conference report or in some combinations of these publications types
and in this respect one cannot give more weight to some of them. However, on
one hand, a research article is often considered more ''stable'' and reliable
than a preprint or conference report, but, on other hand, a preprint, especially
if it is electronic, spreads quite quickly and reaches the audience before a
journal article as a result of which it is a common practice a preprint to be
cited without mentioning its journal version, if any. Further, if an author of a
new already published result continuous to work on it, then it is possible that
he/she will write a review paper, chapter of a book or a whole book that
contains this result and its developments. If this happens, then these
publications are often more cited than the original ones as they normally explain
the result more widely and in connection with other items. In this way, the
last type of publication receive more weight in a form of citations and there
is not a need this to be done by other means. Of course, e.g., publishing a
book or a review paper is considered as authors good achievement but it
may not contain results belonging to its author(s), however, again, there is
not a need to assign to it more weight as the only objective criterion is the
usage of the work by other authors which, in our case, is reflected by its
citations.

	In conclusion, we share the opinion that without knowing the content
of a citing/cited work it should not be assigned different weights to
different types of publications.

\subsection{Quality of publication carrier}
\label{Subsect6.5-1}

	It is a general opinion that it is significant where a particular work is
published. Consider several examples:
    \begin{itemize}
\item
	There are peer review and not peer review journals. The former are
considered as a better place for publication. Besides, some of the journals
in the first group are distinguished at present by their impact factors
(IF)~\cite[Sect.~5.1.1]{Kermarrec_et_al-2007} (see also the discussion of the
IF in~\cite[Sect.~4.4]{Glanzel-2003}); the greater the IF, the better the
journal.

\item
	There are peer review books and ones simply published by their authors
by paying for that. Moreover, the books can be rated by the prestige of the
publishing company that produced them.

\item
	The electronic preprints in the arXiv database with URL
http://arXiv.org are not so valued as their journal versions regardless
that in the most cases both have almost identical content~%
\footnote{~%
	In this respect it must be noted that there are electronic preprints
with world-class excellent results that never appear in other publications.%
}
The reason for this being that the e-prints do not pass real peer review process.

\item
	Nevertheless that a lot of conference reports pass peer judgement, they
are less valued than, e.g., journal articles or chapters in books.
       \end{itemize}

	These examples show that there is needed of a quantitative measure for
comparing the publication carriers of the scientific works. If such a measure
is available, then it can be combined with the citation analysis for making
better conclusions about papers and/or their authors. An example of such a
measure is worked out in~\cite{Levitta_and_Thelwall-2011}. It is  called
''weighted sum indicator''  and is defined by
    \begin{equation}    \label{6.0}
W(c,t) = c A_t + (1 - c)IF
    \end{equation}
where $A_t$ is the number of citations of a work for $t$ years after
publication, $IF$  is the impact factor of the journal in which the work was
published (evaluated for the year of publication), and $c$ is a constant weight
with  $0\le c \le 1$. Unfortunately this indicator is sensible only for works
published as journal articles in journals for which the impact factor rating
is defined.

\subsection{Time dependence}
\label{Subsect6.6}

	A published paper lives as long as people remember (and use) it and
the often they recall about it, the better it is. In the context of this work we
can paraphrase this as: a paper lives  as people cite it and the more citations
it obtains, the more useful it is.  Accepting this point of view, we see that
the time evolutions of the citations of a work or an author should be taken
into account in the citation analysis.

	The time distribution of paper's citations can tell us a lot about the
interest for it. Consider some clear examples.

	(i) After publication a paper receives a lot of citations for a short
period of time, say 3--12 months, and then its citation stops. This
can be interpreted as essential result(s) contained in the paper but it
is likely to be in a rapidly developing field, where exists a flow of new
results which leads to fast redirection of the readers attention.
	(ii)  Suppose a paper receives relatively constant non-zero  citations
for an unit time for a long period, say 40--70 years. This point to fundamental
result(s) in the work that are used extensively for a long time~%
\footnote{~%
Usually this happens with ''good'' bookes/monographs.%
}
	(iii) Other possibility is that the paper's citations per unit time
grows with the time, reaches some maximum, then continuously falls down to zero
and after that the paper gets random citations from time to time. This is
approximately the mean statistical situation in which the paper content
attracts the readers attention for some time, then new works in its field
appear and the attention begins to fall down until the original paper is almost
forgotten, but some scientists  accidentally find something interesting in (\eg
for historical reasons) and cite it.

	This list of possible time distribution of paper's citations can be
continued, but the important for us inference from it is that this distribution
carries information about the usage of the work by its readers. Here is hard to
make general conclusions. However, excluding some rare exceptional works (which
sometimes appear in new editions continuously), the tendency is that the rate
of appearance of new citations decreases with time and tends to zero. The
objective reason for this is the progress of Science as a result of which new
results and publications appear with increasing speed with time. However, there
are exceptions of this observation. They ordinary belong to the works of
recognized and famous scientist. We shall mention here only the name of Sir
Isaac Newton whose works continue to receive citations continuously centuries
after they were first published which point to his great footprint in the
Science heritages.

	The time-depending bibliometric indices~\eref{5.12}--\eref{5.16}
considered in subsection~\ref{SubSubsect5.2.3} adopt (with exception
of~\eref{5.15}) inverse proportionality of paper's age (measured in years),
which is in conformity with the above mentioned tendency. But in these indices
appear the citation speed $c_i/t_i$ of the paper $p_i$ for the whole its age
$t_i$. We may suppose that if $c_i(t)$ are the citations of $p_i$ for time $t$
after its publication, then the speed $\frac{\od c_i(t)}{\od t}$ is more
objective feature of the paper $p_i$. Since $c_i(t)$ is not a continuous
function of $t$, the derivative in $\frac{\od c_i(t)}{\od t}$ is not defined.
For this reason it can be approximated and replaced by
    \begin{equation}    \label{6.1}
\dot{c_i}(t,\tau) := \frac{c_i(t+\tau)-c_i(t)}{\tau}
    \end{equation}
for a fixed time period $\tau$, which is sensible to be set equal to, e.g., one
year.  The graph of this function contains information of interest as for us.
Of course, the speed characterizes only the paper $p_i$, not the whole citation
list of an author. From the speeds $\dot{c_i}(t,\tau)$, $i=1, \dots, n$, can be
constructed different characteristics (''indices'') for a given citation list,
\eg ''Hirsch speed'', defined as the Hirsch index but the role of citations
being replaced by these speeds, arithmetic mean speed,
 $\sum_{i=1}^n\dot{c_i}(t,\tau)/n$, and so on.

	It is likely that instead of the above speeds a more adequate global
characteristic of a citation list is the (global) citation speed
     \begin{equation}    \label{6.2}
\dot{c}(t,\tau) := \frac{c(t+\tau)-c(t)}{\tau}
		  = \sum_{i=1}^n\dot{c_i}(t,\tau)
    \end{equation}
where $c(t):=\sum_{i=1}^n c_i(t,\tau)$ is the number of all citations of the
author's papers at time moment $t$ (for a period of time $\tau$). Again, it is
reasonable to set $\tau$ to one year, but such a choice is based more on
statistical and phycological reasons, than on some abstract arguments.
Local/global extremums of $\dot{c}(t,\tau)$ and other peculiarities of its graph
can suitably be interpreted, but we shall not speculate on this item.

	The speeds~\eref{6.1} and~\eref{6.2} characterize a citation list but
generally not the author to which it belongs. There are analogues that take into
account the individual contribution of the author like:
    \begin{align}    \label{6.3}
\dot{c_i}^{a}(t,\tau) &:=  \dot{c}_i(t,\tau) \cdot w^a_i
\\		    \label{6.4}
\dot{c}^{a}(t,\tau) &:= \sum_{i=1}^n  \dot{c}_i^a(t,\tau) ,
    \end{align}
where $w_i^a$ is the weight of author's contribution to the paper $p_i$, which
we assume to be $w_i^a=1/a_i$, $a_i$ being the number of authors of $p_i$, in a
lack of information about $w_i^a$.

	If $h(t)$ is an author's h-index at a time $t>0$ (usually measured in
years of author career), then in~\cite{Hirsch_index_or_Hirsch_rate} is argued
that the Hirsch\ndash rate (h\ndash rate) given by the average speed $h(t)/t$ is
an good characteristic of the authors and suitable for their comparison..

	In conclusion, the time distribution of the citations of an author's
published works carries an information that should not be neglected in citation
analysis and scientific evolution of that author.

\subsection{Web resources}
\label{Subsect6.7}

	With the development of the computer networks, in particular the Internet
(the Web), the scientists publish documents on them and use such documents on
equal footing with the ones printed on paper. The most obvious example of this
kind for any physicists nowadays is the arXiv database with URL
http://arXiv.org whose documents and the citations to/from them are included in
the free web service Google Scholar with URL http://scholar.google.com/.

	But even the last sentence
reveal one of the major problems with the web resources, \viz their addresses,
\ie where the corresponding files can be found. To be more specific, let us
talk about the uniform resource locator (URL), known as web address, of a
document, say http://arXiv.org/ (this is the main site page), which is like the
physical location of a standard library of paper documents.~%
\footnote{~%
This analogy is quite rough as the web address may remain unchanged while the
physical carrier of the information (or its IP address) changes. Besides, the
particular location of the physical carrier is insignificant.%
}
The problem is that the web address may be changed easily and unpredictably, \eg
the initial address of the arXiv was http://xxx.lanl.gov/, which immediately
leads to disappearence of the connected with it resources at the original (web)
location in the general case. For reliable data resources this should happen
quire rarely and the previous address should be active some time after the
change and should contain proper information; happily, the initial address of
the arXiv database is still active and contains current copy of the arXiv many
years after the address change has occurred.
	So, when citing a web document we must know and quote its web address
but we cannot be sure that the cited address will be valid at a next moment when
one reads this information; the Internet is full with thousands of dead
(hyper)links. This is in contrast to citing paper documents when information
like article from journal~$X$, volume~$y$, pages~$z_1$--$z_2$ will never change.~%
\footnote{~%
This information is constant also for web (on-line) journals but if, e.g., the
journal web site disappears, then all of the journal files will disappear too
if they reside in this site.%
}
Therefore, when citing web documents, we have
to be sure that they have stable and reliable web addresses.

	Other problem with citing web documents is their content. We are used
the paper documents to have constant content after their publication and this
content can be change with subsequent documents making additions, corrections
\etc to the original text. The same procedure works and for web documents but
the problem is that the content of the original and subsequent documents may be
changed at any moment by anyone having suitable knowledge (\eg of web design)
and access to the corresponding server. Of course, this does not happen on
reliable web sites like http://arXiv.org/, but if, for instance, the document
is on the personal web site of an author, then he/she can change anything on
it, in particular if he/she finds an error in some text it is easer to correct
the original document without any announcement than to write a correction with
suitable explanation to the original document. Now the conclusions is that,
when citing web documents we have to be sure that they have constant and
reliable content; possible changes in it should be done only by independent
different subsequent documents or in its revised versions in which case is
supposed that all previous versions are available and unchanged after their
first publication.

	The above problem can be solved if there will be created organization
in which will be archived copies of the cited documents at the moment they are
cited; an example is the web site http://www.webcitation.org/. Such
organizations may play a role similar to ordinary libraries for paper
documents.

	As noted in~\cite[p.~92]{Glanzel-2010}, the major differences
''between print media and the Web is that time plays a different role on the
Web'' and ''the possibility of an almost continuous change of contents on the
Web''. In particular~\cite[p.~93]{Glanzel-2010}: ''Most bibliometric processes
are cumulative since publications (except for the extremely rare cases of
retractions) and citations are irreversible and bibliographic links cannot be
removed if they have once been established. By contrast, the Web is in terms of
both, content and links in permanent change.'' Besides, the web documents are
in general non quality\ndash controlled refereed products as, in principle,
information can be published online by anybody. For these and other reasons the
bibliometric indications have to be applied with some caution to the Web
rescouses (if they are applicable at all in this case) and in general new
measures are needed for the analysis of web linking of documents.

	From the viewpoint of web (hyper-)linking one can look on usual citing
of paper documents as on links between documents. Such an approach can be
considered as an aspect of a kind of social network and suggest the usage of
some web metrics, known as sites statistics indicators, to citation analysis.
For instance, as the downloads of a web document can indicate its usage (this
factor is quite disputable~\cite[Sect.~5.2]{Kermarrec_et_al-2007}), the loans of
documents of an ordinary library can serve the same purpose (with the
corresponding doubts).

	At any rate, at present more and more scientists use and cite web
documents on the same base as they do with other resources. For this reason the
web citations should be count on the same footing as citations in paper
documents.


\section {Citations and scientific achievements}
\label{Sect7}

	Until now we dealt with bibliometric data connected with author's
citation lists. The aim of this section is to tray to make conclusions
concerning author's scientific achievement and impact. Recall, at this stage we
have completely ignored the cited and citing papers contents as such data is
not presented in a citation list.

	So, let us have an author's citation list and some its bibliometric
descriptions. What can we say about the author's contribution in the Science
and can on this base be compared different scientists?

	It is a general opinion that the more citations an author (or a
particular his/her paper) has, the more is his/her (its)  impact. But how big?
The problems seems open from quantitative position.

	The different bibliometric metrics reflect different sides of the
problem. Once these metrics are defined (usually without any a priori
arguments), they are a posteriori confirmed or rejected by gathering statistics
for them. \eg evaluating them for a number of scientists with recognized
administrative and/or scientific positions and comparing the metrics values
with their positions. It seems that no one of the existing single number
metrics describes the scientific achievement of an authors in a satisfactory
way, which is in conformity with the ideas of~\cite{Rosinger-2012} that this
cannot be done in this way.

	A possible measure of an author (or a paper) scientific impact may serve
the time distribution of his/her (or it) citations, in particular his/her (it)
citation life, \ie the period after which citations stop (which does not mean
that they will not appear in future). In general, the longer the citation life,
the bigger is the scientific impact. But how big? The problem seems open from
quantitative point of view. Also, it seems that the more uniform/even and large
the distribution is, the bigger the impact may be but, again, a quantitative
measure is missing. Besides, the peaks in the distribution point to a temporary
rasing interest to an author or his/her paper(s).

	Obviously, there is a connection between citations and their
bibliometric measures with the scientific impact of an author or particular
his/her paper(s). At present this connection is far from being well
investigated and the known results in this field of research are mainly based
on statistical analysis~\cite{h-index_and_9_its_variants,
h-index_and_37_its_variants}. By the last we mean that after some bibliometric
measure (index, metric) is chosen, it is calculated for a selection of
scientists (\eg from departments of an institute/university, for winners of
some prize), then the results are compared with the known administrative and/or
scientific positions of the scientists and, finally, on this base are drawn
conclusions for the adequateness, in particular the pros and cons, of the
measure chosen.~%
\footnote{~%
In this connection we want to point to the
paper~\cite{h-index_for_147_chemistry_research_groups} in which the h-index
is calculated for 147 university chemistry research groups and the results are
compared with standard bibliometric measures and of peer review
judgment and a correlation in this respect is observed.%
}
If the measure receives sufficient number of pros in some
field, then it is accepted in this field, but it is often applied to other
fields which sometimes shows inadequate results. Exactly this is the case with
the Hirsch and Hirsch\ndash like indices when they are used for evaluation of
authors scientific impact or for comparing scientists.

	Let us say a few words on the total number of author's published works
and their total number of citations. The total number of published works is a
measure of author's productivity, not of his/her scientific impact. In this
respect a more adequate measure is the coefficient of citation
performance~\eref{5.21} or the published works with non\ndash zero citations.
Besides, to be more precise, one should count only the author's contribution,
in which case the individual effectiveness~\eref{5.21-1} and the individual
publications with non\ndash zero  citations given via~\eref{5.21-3} should
be considered.~%
\footnote{~
contribution, then the number $n$ of all his/her publications should be
replaced by the effetive/individual number of publications~\eref{5.21-2}.%
}


\section {Taking into account papers content}
\label{Sect8}

	For pure bibliometric purposes the content of citing and cited works
does not matter. But when one begins to interpret and use bibliometrics for
scientific evaluation of authors and their works, the content begins playing
essential role and in a lot of cases it is more important than citation metrics.
Unfortunately, in this field the problems are more than the solutions.

	Suppose in a citing paper is said and proved that the cited one contains
plagiarism(s) and nothing more. For the bibliometrics this simply adds one more
citation for the author(s) of the cited work but the common sense tells us that
here is something terribly wrong. Our suggestions is to count such citation
(if the stated in them is true of course) with negative sign, \ie by
subtracting their number from the other author's citations.

	Next, let a citing paper points to error(s), wrong result(s), \etc
in the cited work. Now the situation is not so simple as it may look and the
assessment depends on (subsequent) details. If it is said that the cited work is
so wrong that it cannot be corrected (\eg a theory contradicting to
experiments), then such a citations are reasonably to be neglected, \ie they
have not to be counted in the citations list. Contrary to this possibility, if
the wrong result(s) are not only pointed but also corrected and then strictly
proved in the citing paper, then this means that the cited work has influence
on the citing one with inspiration of finding new result(s) in which case the
citation may be treated as an ordinary one. There are many other possibilities,
like simple pointing to error(s) in which case the citation may be omitted,
but we do not want to speculate on them.

	There are also ''neutral'' citations in which nothing is said about a
cited paper. Examples are simple lists of works on some item(s), mentioning of
the cited paper in introduction/conclusion in connection with some problem or
in a general list of references on the subject of the citing paper. The purpose
of such a citation is to point readers attention to the cited paper without
giving opinion on it. This sort of citations are completely covered by the
bibliometrics and should be count as ordinary citations.

	Consider now the most difficult problem when a citing work makes
particular use (of part(s)) of the cited one. This is the most creative reason
for citation as the cited paper has directly influence the citing one in a
positive way. It is intuitively clear that to such citations is fair to be
given more weight than, e.g., the neutral ones. This weight should surely depend
in the particular usage of the cited paper but the problem  for its
quantitative measure is open. Let us mention some possibilities:
    \begin{itemize}
\itemsep 4pt
\renewcommand{\labelitemi}{$\diamond$}
\item following/developing method(s) introduce in the cited work
\item application of particular result(s) from the cited work
\item testing result(s) from the cited work for particular events/data
\item using result(s)/idea(s) from the cited work as a ground for further
research .
    \end{itemize}
It is quite obvious, any one of these and many more situations gives arguments
for assigning to such kind of citations greater weight than \eg ordinary
citations or negative citations (\eg revealing a plagiarism). But, as we
already said, a quantitative measure for such weight is missing.

	In subsection~\ref{Subsect6.5} we have presented arguments that the
different types of publications should not be distinguished for pure
bibliometrics reasons. But what happens if the content of the works is taken
into account?

	The main output of a research paper are new results, ideas, concepts,
methods, \etc and these are the main reasons for citing them.

	The main output of (scientific) popular works are presentations of
known and established knowledge in a way that can be understand by wider range
of people, \eg non\ndash scientists or scientists from other fields of
research. So, these papers can be regarded as review works written for
non\ndash specialists in the material they cover.

	The general purpose of a (scientific) review paper is a systematic
detailed presentation of material from published research papers, usually on
some fixed topic. Often review papers give unified notation system, present
and compare different aproaches/ideas, contain proofs, discuss pros and cons
and are easier to read than the original works they cite.  These and other
features attract more readers and in this way they contribute to the spread and
acceptence/non\ndash acceptance of the information from the papers they review.

	The books/monographs are considered as the most ''heavy'' and reliable
sort of publications.~%
\footnote{~%
We exclude from our considerations the ''self\ndash published'' and similar
books (or other publications) when an author pays a company to publish the
books without any realistic peer review process. In this connection it should be
mentioned that the reputation and respect of the publishing company among the
scientists is also essential.%
}
Normally they are the most detailed and different\ndash sided presentations of
the topic(s) they cover and contain suitable references which in some cases are
quite intensive. For these and other reasons a good book can be used by other
authors for many years by putting aside review and original works on its
topic(s).

	The textbooks are books written for educational purposes and hence
usually presuppose less preliminary knowledge compared to monographs on the
same subject. Besides, they normally include material that is accepted with
certainty as a true by the scientific community and only partially concern
latest scientific news. However, often good textbooks can be regarded as
monographs and vice versa. The textbooks may be cited more rarely than
monographs, but often they leave a greater footprint on their readers by giving
them  basic background for further research/development.

	We can continue to list and partially analyze other publication types
like handbook, encyclopedia, chapter in book/collecton and so on, but the above
material is enough to confirm the opinion that to different publication types
may be assigned different weights. However, a quantitative way for doing this is
not known.

	Until this section we talked about external aspects of the citation
process with respect to the cited papers. The most important thing  of any
paper is its content. Its citation and all connected with a paper ratings and
scientific impact are consequence of its content.  But the content of any
particular paper is specific and, besides, its evaluation strongly depends on
the particular readers of the paper (\eg of their education, knowledge and even
phycological conditions).~%
\footnote{~%
Other problem is how a published work finds its way to its readers.%
}
The final decision on author (resp.\ paper) scientific impact and value
is formed by the scientific community on the base of the opinions of persons
acquainted with the content of his/her published works (resp.\ the given paper).
At present is not known a  way to formalize this human\ndash dependent process.~%
\footnote{~%
We do not want to speculate how such a process can be manipulated. The history
reveals that manipulations in this field are short\ndash lasting and after some
time the scientific truth takes its place. It is enough to recall here the
church manipulations of the famous investigations of Galileo Galilei.%
}

	For these and many other reasons it is a great responsibility of the
pears to evaluate the scientific results, impact, achievements and position
of a person. In particular this concerns the decisions for giving prises,
honorable and scientific degrees, \etc


\section {Implicit citations or citations without citing}
\label{Sect9}

	When evaluating the achievements of a scientist there are also other
objective criterions than the ones based on citations.

	Possibly the most important and recognized tribute to a scientists is
by directly connecting his/her name with some scientific  formulation like
experiment, law, constant, equation/formula, observation, idea, hypothesis,
theory, etc. There are thousands of such examples, for instance:
Mikelson\ndash Morley experiment, Newton's (first, second, third, gravity) law,
Boltzman constant, Schr\"odinger equation, Galilei moons, M\"ossbouer effect,
Dirac large number hypothesis and Einstein theory (\eg of gravitation). In this
way we not only recognize  the scientific impact of a scientists but also pay
tribute to his/her personal work and role in the Science. In this way when we
say/write, e.g., Plank constant we implicitly have in mind the contribution(s)
of Max Plank to quantum physics in the particular case and there is not need to
cite his paper(s) on early  quantum physics which any one familiar with Plank
constant can find and cite easily. In this sense here we have an implicit
citation of scientist's work(s) whose weight is certainly more than a simple
citation of a particular his/her paper, but a quantitative measure of this
weight is missing.

	Other way of scientific impact is via symbols, notations, concepts,
names of different results/objects introduced by known scientists(s) which do
not have his/her (their) name(s) in written version. Examples are the plus sign
''$+$'' (the teacher Michael Stifel, 1544), the speed of light (in vacuum)
constant $c$ (W.~E.~Weber and R.~Kohlrausch, 1858), the ''equivalence
principle'' in gravity theories (A.~Einstein 1907, but the origin can be seen
in G.~Galilei experiment demonstrating the independence of the gravity
acceleration of bodies in vacuum from their masses), the ''classical
electromagnetism'' theory (in its present day understanding)  due mainly to
J.~C.~Maxwell (but behind this theory stay also many other scientists),
Special/General theory of relativity (A.~Einstein, 1905/1916), and ''isospin''
(W.~Heisenberg 1912). By using such notations, names, concepts \etc we accept
their importance and role in the Science and thus implicitly recognize the
contributions of their inventors which can also be considered as an implicit
citation of their works.

	Behind many physics titles like kinetic gas theory, standard model (of
elementary particles), quantum mechanics and classical/quantum field theory stay
the work of many scientists. A check in the corresponding (historical)
resources reveals to whom they are due to. Thus the usage of these titles
(even without any explanations) is an implicit way to pay tribute to the
persons that have contributed to their formulation and the physics content that
they have.

	When citing \eg books, textbooks and review papers one often has in
mind results, experiments, formulations \etc which do not belong to the
author(s) of these publications and are only collected, summarized or
reformulated by him/her (them). In this sense such citations pay tribute not only
to the work of the direct author(s) of the publication but also indirectly to
the persons who have made, for instance, the discovery describe in the cited
paper. Namely the collection of a lot of material that generally does not
belong to the authors of review works is one of the reasons that makes them
convenient to cite without mentioning the original resources, which, on other
hand, brings more citations to the review papers.


\section {Peer judgements}
\label{Sect10}

	The written in the last two sections shows explicitly that when
evaluating the impact of scientific works and their authors there are important
factors that should be taken into account and that are completely out of the
range of the bibliometrics. At present these factors are only in
the range of the peers. The peers are qualified experts in some field of
Science that give their opinions on some scientific works in this field and
their authors and possibly give some recommendations about them. The peers are
supposed to know very well the considered scientific field and to be able to
write corresponding reports (peer reviews) on papers in it in which they
evaluate the scientific research for originality, competence, significance,
\etc The peer reviews play roles like advice system, quality control and error
detection. On the base of the peer reports are taken further decisions like
acceptens/non\ndash acceptance of papers for publication, promotions, ratings
of works and/or their authors, \etc These reports are of crucial importance for
non\ndash experts which form their opinions on this ground; in particular in
this way is generally formed the public opinion on the scientists. The peer
reviews are central for many key problems of Science like quality control and
decision-making.

	The work~\cite{Bornmann-2011} is a comprehensive review of the peer
review process: ground and background, purposes, advantages and disadvantages,
problems, perspectives, reliability and fairness. It also contain a detailed
bibliography on the item.

	Of course, the peer judgements are human-depending activities and,
respectively, may be influenced by non\ndash objective reasons like the
phycological condition of the particular peer, his/her personal relations with
some author(s)/scientist(s), his/her interests and knowledge of the field of a
reviewed material, \etc Such factors are partially remove by taking into
account reviews written independently by more than one peer (usually 2 or 3
peers), but the personal elements of all of them remain and it is up to other
persons, for example (managing) editors or super-peers, to try to eliminate them
or/and to make more objective decisions.

	Until now the peer judgements are not formalized in a form of some
algorithms and it is unlikely that this will ever be done. The bibliometrics
provides quantitative methods for analysis of the scientific and technological
literature and in this sense it helps for revealing the impact of the
scientific papers and their authors. It should be noted that a lot of these
methods are based on statistical data analysis as a consequence of which the
results point to tendencies or/and statistical laws the automatic application
of which to particular papers/authors may lead to wrong conclusions. One of the
roles of the peers is to decide upon the applicability of these results to
particular situations. On the opposite, the bibliometric results may be used to
trace statistically the validity of the peer reports. All this points that peer
reviews and the bibliometric evaluations should be regarded as complimentary to
each other and used simultaneously for obtaining better assessments of the
scientists and their papers.


\section {Conclusion}
\label{Conclusion}

 The evaluation of the scientific value and impact of the works of a scientist
is important for many purposes, in particular for comparing with other
scientists, promotion and recruitment, prize awarding, fellowship, and funding.
The main methods for such an evaluation are peer review judgments, based
on the opinions of group of experts, and citations analysis. Of course, a
combination of the both methods is possible and seems a better choice.

In the paper~\cite{h-index_for_147_chemistry_research_groups} is calculated
the h-index for 147 university chemistry research groups and the results are
compared with standard bibliometric measures and of peer review judgment. It is
important that a correlation in this respect is observed which, in particular,
 means that the h\ndash index follows in general the peer judgment.

In~\cite{Costas_and_Bordons-2007} is analyzed the h\ndash index in different
situations and its relations with standard bibliometric characteristics like
total number of the papers and their total number of citations, citations
per paper, highly cited papers (with no less than 15 citations), impact
factor(s) based on impact factors of the journals in which an author has
published, etc.\ The general conclusion is that the Hirsch index is a
good thing but it alone cannot be a ''complete'' measure of a scientists and it
should be complimentary to other bibliometric measures. To overcome its
disadvantages were introduced many other indices each of which has its pros and
cons~\cite{Alonso_et_al-2009} but no one of them cannot pretend to be an
ultimate measure of an author impact.

	In this respect we want to note that the Hirsch index is not adequate
when some or all of the papers in its core have more than one author as it
assigns all of the work of these authors to one of them and, respectively, the
achievements of this work are also assigned to the author whose citation list is
considered. The senseless of this situation is evident if we take an  $n$\ndash
author paper with $n\ge2$ and calculate the Hirsch index for any one of the
authors of this paper. If it falls into the core of the Hirsch index for all
authors, then any one of them can claim the "fame" of this
paper belongs to him/her which will mean that the whole "fame" of the paper is
$n\times100\%$ instead of 100\%.

	Let us note that the automatic calculation of bibliometric indices based
on Internet databases generally depends on the database~\cite{Bar-Ilan-2008}.
Besides, the Internet databases do not capture all existing
citations~\cite{Nisonger-2004}. In general the results of the application  of
the bibliometrics depends on the used data sets and the captured publications
(which is always limited). Besides, bibliometrics cannot measure procedures like
reviewing, editing and mentoring. In this sense it has serious limitations.
Similarly, the peer reviews have limitations too~%
\footnote{~%
 	An excellent review on the process and results of peer review research
is presented in~\cite{Bornmann-2011}.%
},
but each of the both methods partially corrects the disadvantages of the other
one.
  	It is observed a correlation between assessment by different
bibliometric indicators and quality judgment of
peers~\cite{h-index_for_147_chemistry_research_groups, Waltman_et_al-2011,
Bornmann-2011}. This naturally suggest that~\cite[p.~229]{Moed-2005} the
bibliometrics can be used as a supplementary tool in peer review process as well
as either of them can be used as validation and monitoring tool relative to the
other one.

   	Besides documented via citations usage of published works, there can
be a lot of other their usages that are not recorded, \eg full or partial
viewing/reading without citing, hearing about them on a seminar, conference or
a private conversation \etc It is practically impossible to count and/or
measure such events and to evaluate their impact but it is clear that they are
due to the authors of the discussed papers and in this sense they contribute
to the authors fame and impact on other scientists.

	As noted in~\cite[Sect.~5]{Kumar-2009} the gathering of page/site
statistics of Internet pages with author works can be used for conclusions
about the author. In fact, when certain web pages are a (partial) home of
authors works like abstracts, partial/full text papers, lists of titles
(possibly with further links), files with data etc., then from the statistics
of such pages~%
\footnote{~%
Usually a web page/site statistics includes data like number of (unique)
visitors, number of visited pages, number of downloaded files, the distribution
of these numbers in time, as well as more detailed info like the particular
visited pages and the time spend on them, list of downloaded files,
user-dependent data (IP address, country, browser, \etc), and many more. %
}
can be made different conclusions. For instance, the page views and downloads
of author works files can be interpreted as an interest of other people about
the particular author and his/her works. Of course, from these raw data cannot
be made unique conclusions, \eg the fact that someone has viewed a
particular page for one hour does not mean that he/she has read this page for
one hour as he/she may simply doing other things and forgot to leave the
particular page. However, in~\cite[Sect.~5]{Kumar-2009} is reported a
particular example when ''high viewership does lead to high citations, and
highly cited articles do not necessarily have high viewership''.

	In this respect as a Web analogue of the standard citations may
be considered the Internet (hyper)links to web pages that contain authors
papers or/and relevant information about them. Such pages may be from author's
personal web site, databases with works like the e-print archive
http://arxiv.org/, publisher or journal web site, and so on. Regardless that
such links can be generated automatically by robots, from them can be made
conclusions similar to the ones from the citations. It is clear that
such an approach to author impact is in favour of authors that (extensively)
use the capabilities of Internet but such are the present day realities and
possibilities. As an argument in favour of such measure may serve the reported
in~\cite{Vaughan1-Shaw-2005} statistical result that ''Web citation correlates
with ISI citation and the average Web citation count of a journal correlates
with the Journal's Impact Factor'' in biology and genetics; besides ''Web
citations show a broader geographic coverage and capture a greater number and
variety of users of journal articles''. However, the fact that Internet links
and documents are unstable is a big problem. By "unstable" we understand that
they may change within seconds or simply become not valid; \eg a web page may
disappear alone or with a part of the site that contains it and the content of
a web page may be changed at any moment from the corresponding web
designer/administrator. For this reason it is a good idea to be made a copy of
a web resource when citing it as a proof for its existence and content at the
moment when it was used.~%
\footnote{~%
But the things do not end here as the copies of web pages can be manipulate
extremely easy by anyone knowing some web design. As a real proof of the
existence and the content of a web page (or a link) its copy should be archived
in a publicly available and respected place the documents in which should have
the same reputation and reliability as the paper books in an ordinary public
library.%
}

	The modern Science is due to a great extend to the research and its
assessment by peers. The methods and tools of bibliometrics are an alternative
to the work of peers. Since aspects such as the quality and impact of a paper
are not yet formalised in a strict mathematical sense, the peer reviews remain
leading in the final decisions on these items, but the bibliometric indicators
reveal some objective their properties and tenencies. The both approaches seem
to be overlapping and complimentary to each other which stimulate the further
development of strict methods for assessment of papers and their authors.


\section*{Acknowledgments}

	The author thanks to Prof. Pavel Stavrev for renewing his interest in
the problems for strict evaluations and impact of scientific publications and
for numerous discussions on this item.



\addcontentsline{toc}{section}{References}
\bibliography{Bozho-BiBTeX-Refs-Bibliometrics}
\bibliographystyle{unsrt}
\addcontentsline{toc}{subsubsection}{This article ends at page}




\end{document}